\documentclass[11pt]{article}
\usepackage[margin=1in]{geometry}
\usepackage{authblk}
\usepackage[symbol*]{footmisc}
\setfnsymbol{wiley}

\usepackage[table,dvipsnames]{xcolor}
\definecolor{NCSUred}{RGB}{204,0,0}
\definecolor{NCSUyellow}{RGB}{250,200,0}
\definecolor{NCSUorange}{RGB}{209,73,5}
\definecolor{NCSUdarkred}{RGB}{153,0,0}
\definecolor{NCSUaqua}{RGB}{0,132,115}
\definecolor{NCSUgreen}{RGB}{111,125,28}
\definecolor{NCSUblue}{RGB}{65,86,161}
\definecolor{NCSUlightblue}{RGB}{66,126,147} 
\usepackage{graphicx, url}
\usepackage[colorlinks=true, linkcolor=NCSUblue, urlcolor=NCSUaqua, citecolor=NCSUblue]{hyperref}
\usepackage{amsmath, amsfonts, amssymb, amsthm}
\allowdisplaybreaks
\theoremstyle{plain}

\RequirePackage[T1]{fontenc}
\usepackage{stix2}
\usepackage[cal=rsfso, bb=ams]{mathalpha}

\newcommand{\mc}[1]{\mathcal{#1}}

\newcommand{\mr}[1]{\mathrm{#1}}

\newcommand{\mR}{\mathbb{R}}

\newcommand{\mN}{\mathbb{N}}
\newcommand{\xD}[1]{\mr{d} #1}
\newcommand{\xDD}[2]{\frac{\xD{#1}}{\xD{#2}}}

\newcommand{\ip}[2]{\langle #1, \, #2 \rangle}

\title{\bf Convex Hybrid Modeling: An Operator-Based Approach\thanks{A 6-page shortened version under the same title is submitted to 2027 \textit{Foundations of Computer Aided Process Operations (FOCAPO) / Chemical Process Control (CPC) Conference}. This is the full-length version.}}
\author[1]{Wentao Tang \thanks{Corresponding author: \href{mailto:wtang23@ncsu.edu}{\tt wtang23@ncsu.edu} }}
\affil[1]{Department of Chemical and Biomolecular Engineering, North Carolina State University}

\date{May 21, 2026}

\begin{document}
\pagenumbering{roman}
\maketitle
\pagenumbering{arabic}

\begin{abstract}
    While machine learning can accurately model process systems, models for decision making should also be structurally simple and physically interpretable. In process control, for example, (nearly) linear models are favored than nonlinear ones, promoting the use of operator theory, which ``universally'' represents a nonlinear system by a nonparametric operator. On the other hand, interpretability requires by a ``non-universal'', parametric nonlinear model family satisfying first principles; these constraints tend to complicate the learning procedure. This paper considers hybrid modeling by formulating convex learning problems that account for interpretability systematically and give surrogate models efficiently. Three settings are discussed -- (i) regularization around a particular ``reference model'', (ii) restriction on an ``interpretable subspace'', and more generally, (iii) restriction on a ``interpretable manifold'' that is nonlinearly parameterized. In the more general setting, by introducing an operator-theoretic technique to re-parameterize models in the ``lifted'' parameters (``canonical features'', potentially infinite-dimensional), the system is regarded as a kernel-based mixture of interpretable models. Application to both static and dynamic models are exemplified in numerical studies.
\end{abstract}

\section{Introduction}

\par While machine learning-based (especially deep learning-based) surrogate models have become well-received in process systems engineering, the concern with their lack of interpretability (and hence consistency with physical first principles) motivated researchers to consider the problem of \emph{hybrid modeling} \cite{yang2020hybrid, chakraborty2022hybrid, mitrai2025accelerating}. This requires a combination of data-driven learning and mechanistic modeling -- an idea that existed for more than three decades \cite{psichogios1992hybrid, su1993integrating}. 
In Bradley et al. \cite{bradley2022perspectives}, regarding the relation between first-principles and data-driven components in a hybrid model, six strategies were summarized: (a) emulation, (b) assimilation, (c) estimation, (d) physics-informed learning, (e) structuring, and (f) calibration. 
In Chakraborty et al. \cite{chakraborty2022hybrid}, regarding the representation of firstprinciples in a hybrid model, six pathways were listed: (a) feature engineering, (b) customized knowledge representation, (c) additional constraints, (d) integration of these approaches, (e) custom model structure, and (f) end-to-end domain-specific AI models. 
A tutorial on hybrid modeling with a review on a variety of successful applications was given in Shah et al. \cite{shah2025hybrid}. One can easily note that in the majority of these works, the hybrid models are based on (deep) neural networks. 

\par If the hybrid models are intended for decision making -- process design, control, and operations -- then the quality of these models should have not only satisfactory prediction accuracy, but also a mathematical structure that is \emph{simple and useful for reliable decision making}. 
For example, for control, one can argue that linear dynamical models are preferred over nonlinear models, since the controller synthesis or optimal control problems associated with linear dynamics are usually formulated as convex optimization problems and can be computed in a fast and reliable way. 
This explains the increasing popularity of \emph{Koopman operator theory} in today's control theory development. Specifically, for a nonlinear autonomous system in discrete time: $x_{t+1}=f(x_t)$, the Koopman operator refers to the linear operator on a function space, defined by $K: g\mapsto g\circ f$, thus representing the nonlinear dynamics as an infinite-dimensional linear one. Instead of identifying the nonlinear state transition map $f$, one can more efficiently learn and examine the properties of the operator $K$ \cite{bevanda2021koopman, kostic2022learning, shi2026koopman}. 
Recently, many nonlinear control problems (such as stability analysis, dissipativity analysis, and state observer synthesis) have been formulated in linearized, operator-based forms and solved in a data-driven manner \cite{tang-ye2025koopman, tang-ye2026koopman, tang-ye2026dissipativity}. 
\footnote{Even for models of nonlinear structures such as neural networks, optimization over such nonlinear models typically adopt a perspective of globally linearizing them on an infinite-dimensional function space, e.g., by considering neural networks of infinite/indefinite width and depth as members of reproducing kernel Banach spaces or by considering optimization in a space of probability measures \cite{jacot2018neural, bartolucci2023understanding, chen2025accelerating}. The mechanistic understanding of noncovnex training procedures in deep learning, so far, largely remains an open question.} 

\par Roughly speaking, if one is only concerned with the problem of model identification from data, then from a computational efficiency and expressiveness point of view, it is advantageous to adopt a \emph{convex} learning approach over a rich or even nonparametric family of models. For example, to learn a static model function $h: \mathbb{X}\rightarrow \mathbb{R}$ within a reproducing kernel Hilbert space (RKHS) $\mc{H}_\kappa(\mathbb{X})$ specified by a kernel function $\kappa$, the least-squares or regularized least-squares problem, known as the kernel regression problem, is convex and in fact can reduce to a finite-dimensional subspace, i.e., $h(\cdot) = \sum_{i=1}^n c_i\kappa(x_i, \cdot)$. 
On the other hand, if interpretability or physical first principles are also required to be met, then they must be imposed as hard constraints, extra loss terms, or a family of nonlinearly parameterized models: $\{h(\cdot;\theta): \theta\in \Theta\}$, These may not preserve convexity and tend to complicate the learning problem. 
If neural networks or other nonlinearly parameterized surrogate models are used (which anyways gives nonconvex learning), then such complication does not \emph{further worsen} the nonconvexity. Yet, when a linearly structured model is desirable, then it is imperative that the pursuit of interpretability does not destroy the intended tractability.

\par Hence, this paper focuses on the problem of \emph{convex hybrid modeling}, i.e., that of reconciliation between machine learning (towards an expressive, accurate model) and first principles (towards an interpretable, physically meaningful model), without losing convexity in the optimization formulation or linearity of the resulting model. 
To this end, we propose that the enforcement of interpretability can be usually performed in three paths: (i) regularization around an explainable reference model that already satisfies physical principles, (ii) restriction on a linear subspace of the total model class, where the subspace is considered as the subclass of models consistent with first principles, and (iii) restriction on a \emph{nonlinear manifold} of interpretable models, with a general nonlinear parameterization structure. 
The last setting is viewed as the most general and common one in process systems. 

\par To tackle the nonlinear parameterization in a convex optimization framework, thus formally reducing case (iii) to (ii), operator-theoretic notions are used. 
Specifically, we introduce a ``lifting'' of the nonlinearly parameterized model family into a \emph{reproducing kernel Hilbert space} (RKHS), assuming regularity of the model functions. 
In the RKHS formulation, a model is a bilinear mapping from the Kronecker product between the canonical feature of the input variables $\phi(x) = \kappa(x, \cdot)$ and the canonical feature of parameters $\psi(\theta) = \varkappa(\theta, \cdot)$. 
Thus we regard an interpretable model as a ``kernel-based stochastic mixture'' that reside in an (usually infinite-dimensional) simplex of probability distributions over model parameters. As such, the hybrid model learning problem preserves convexity, and is computationally tractable in a data-driven solution (based on a sampling of the parameter space). 
As we will see, the approach is applicable to both static modeling and dynamic modeling. 
Representative examples from classical engineering thermodynamics and nonlinear process control are used to demonstrate the proposed approach.

\section{Setting I: Reference Model}
We first consider the simplest setting for hybrid modeling, where the physical first principles are taken into account by a fixed model $h^\circ$ that is ``interpretable''. The model to be identified is another function $h$ that deviates from $h^\circ$ to achieve better model accuracy. 
To reconcile the interpretability and accuracy, one then requires that the deviation $h - h^\circ$, namely the \emph{residual} model, be regulated. 

\subsection{Formulation as Kernel Ridge Regression}
\par For a meaningful formulation, we assume that $h^\circ: \mathbb{X}\rightarrow \mR$ belongs to a Hilbert space $\mc{H}$, a potentially infinite-dimensional function space endowed with an inner product $\ip{\cdot}{\cdot}$, for example, $L^2(\mathbb{X})$, the space of square-integrable functions, on which $\ip{h_1}{h_2} = \int_{\mathbb{X}} h_1(x)h_2(x)\xD{x}$ is the inner product. 
A better choice that enables theoretic guarantee of generalization errors and efficient computation is a \emph{reproducing kernel Hilbert space} (RKHS) \cite{cucker2002mathematical, cucker2007learning}. The class of functions lying on the RKHS is a much smaller class than $L^2(\mathbb{X})$ -- roughly speaking, it only contains the functions that have the same ``regularity'' specified by its kernel function $\kappa$ \cite{steinwart2008support}. 
Specifically, if $\kappa: \mathbb{X}\times \mathbb{X}\to \mR$ is a symmetric bivariate function such that for any finite number of points $x_1, \cdots, x_n\in \mathbb{X}$, the matrix formed by $\kappa(x_i, x_j)$ (known as the kernel matrix or Gram matrix) is positive semidefinite, then the space 
$$\mc{H}_\kappa(\mathbb{X)} = \mc{H} := \overline{\mr{span}}\left\{ \sum_{i=1}^n c_i\kappa(x_i, \cdot): 
\begin{matrix}
    x_1, \cdots, x_n\in \mathbb{X}, \\ c_1, \cdots, c_n\in \mR, n\in \mN
\end{matrix} \right\}$$
is a Hilbert space. On this Hilbert space, namely the RKHS specified by kernel $\kappa$, the inner product is given by $\ip{\kappa(x, \cdot)}{\kappa(x', \cdot)} =\kappa(x,x')$ and satisfies the so-called reproducing property: $\ip{h}{\kappa(x,\cdot)} = h(x)$ ($\forall h\in \mc{H}$ and $x\in \mathbb{X}$). 

\par On the RKHS, the magnitude of a function is thus naturally characterized by its RKHS-norm induced by the inner product on the RKHS. 
Hence, given a data sample $\{(x_i, y_i)\}_{i=1}^n$ from the unknown population over $\mathbb{X}\times \mathbb{Y} \subset \mathbb{X}\times \mR$, a hybrid model is sought by solving the following regression problem over the Hilbert space $\mc{H}$:
\begin{equation}\label{eq:regression.1}
    \min_{h\in \mc{H}} \enspace \sum_{i=1}^n \left( h(x_i)-y_i\right)^2 + \lambda \left\|h-h^\circ \right\|_{\mc{H}}^2. 
\end{equation}
Here $\lambda>0$ is a regularization parameter. At the small $\lambda$ limit, the regression problem \eqref{eq:regression.1} reduces to a simple least squares problem, and in fact, due to the infinite dimension of $\mc{H}$, when $\mc{H}$ is an RKHS, reduces to a kernel interpolation problem. At the large $\lambda$ limit, the solution to the regression problem is pinned onto $h^\circ$, with data sample playing a negligible role. Hence $\lambda$ determines a tradeoff between modeling from first principles and learning from data. 
Rewriting the residual model $h-h^\circ = r\in \mc{H}$, and denoting $\eta_i = y_i - h^\circ(x_i)$, the problem becomes a \emph{kernel ridge regression} (KRR) problem on the residual model:
\begin{equation}\label{eq:regression.1.KRR}
    \min_{r\in \mc{H}} \enspace \sum_{i=1}^n (r(x_i)-\eta_i)^2 + \lambda \|r\|_{\mc{H}}^2. 
\end{equation}

\par This is an infinite-dimensional convex optimization problem. Nevertheless, on an RKHS, the well-known \emph{representer theorem} states that the above optimization problem admits an optimal solution for function $r$ on the $n$-dimensional subspace $\mc{H}_n = \mr{span}\{\kappa(x_i, \cdot): i=1,\cdots,n\}$. 
This is because for the objective functional, the first term can be written as the average of $\ip{r}{\kappa(x_i, \cdot)}^2$. If decomposing $r$ as $r_\parallel+r_\perp$ with $r_\parallel \in \mc{H}_n$ and $r_\perp\perp \mc{H}_n$, then $\ip{r}{\kappa(x_i, \cdot)}^2 = \ip{r_\parallel}{\kappa(x_i, \cdot)}^2$, independent of the complementary component. The second term is clearly $\|r_\parallel\|^2 + \|r_\perp\|^2$, where the complementary component $r_\perp$ only adds to the objective. 
By letting $r=\sum_{i=1}^n c_i\kappa(x_i, \cdot)$ and simplifying the objective to a quadratic form of $c=(c_1, \cdots, c_n)\in \mR^n$, \eqref{eq:regression.1.KRR} admits an explicit solution:
$$c^* = \left(G + \lambda I \right)^{-1}y, $$
with $y=(y_1, \cdots, y_n)$ and $G = \left\{\kappa(x_i,x_j) \right\}_{i,j=1}^n$ being the Gram matrix. Hence, the determined hybrid model is 
$$h(x) = h^\circ(x)+\sum_{i=1}^n c^*_i\kappa(x_i, x). $$
The acquisition of the best hybrid model then follows from the fine-tuning of the regularization parameter $\lambda$, e.g., by grid search and cross-validation. 

\subsection{Example: Modeling Phase Equilibrium around a Relative Volatility Model}
\begin{figure}[!t]
  \centering
  \includegraphics[width=\columnwidth]{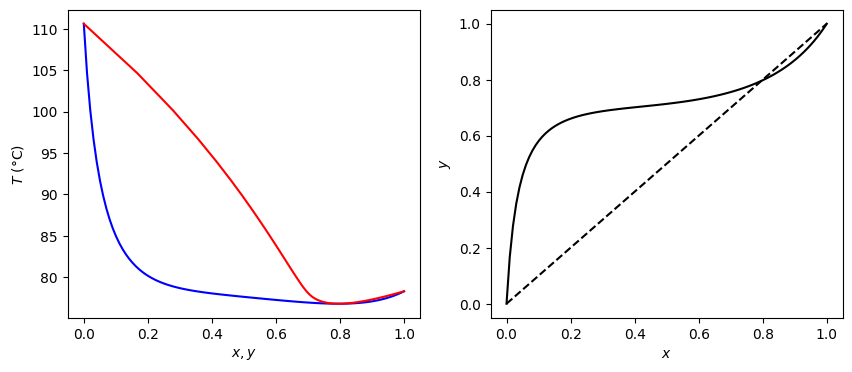}
  \caption{Phase diagram of the ethanol--toluene mixture. ($x$, $y$: molar fraction of ethanol in the liquid and vapor phases, respectively.)}
  \label{fig:phase_diagram}
\end{figure}

The calculation of vapor--liquid equilibrium is useful for designing the separation processes of binary mixtures. Here we consider the ethanol--toluene mixture, which is known to form an azeotrope (with a boiling temperature of about $76.7^\circ\text{C}$), and hence has a nontrivial thermodynamic model. We take the UNIQUAC model as the ground truth, from which the data is collected. The activity coefficients in a UNIQUAC model is given by 
$$
\begin{aligned}
    \ln \gamma_1 =& \ln\frac{\varphi_1}{x_1}+ 1-\frac{\varphi_1}{x_1} - 5q_1 \left( \ln\frac{\varphi_1}{\vartheta_1}+ 1-\frac{\varphi_1}{\vartheta_1} \right) \\
    &+ q_1 \left( 1 - \ln(\vartheta_1+\vartheta_2\tau_{21}) - \frac{\vartheta_1}{\vartheta_1+\vartheta_2\tau_{21}} - \frac{\vartheta_2\tau_{12}}{\vartheta_1\tau_{12}+\vartheta_2} \right) ; \\
    \ln \gamma_2 =& \ln\frac{\varphi_2}{x_2}+ 1-\frac{\varphi_2}{x_2} - 5q_2 \left(\ln\frac{\varphi_2}{\vartheta_2}+ 1-\frac{\varphi_2}{\vartheta_2}\right)  \\
    &+ q_2\left( 1 - \ln(\vartheta_1\tau_{12}+\vartheta_2) - \frac{\vartheta_1\tau_{21}}{\vartheta_1+\vartheta_2\tau_{21}} - \frac{\vartheta_2}{\vartheta_1\tau_{12}+\vartheta_2}\right) . \\
\end{aligned}
$$
Here $\varphi_j = x_jr_j/\sum_i x_ir_i$, $\vartheta_j=x_jq_j/\sum_i x_iq_i$, and $\tau_{ij} = \exp(-a_{ij}/T)$. The parameters are $r_1=2.1055$, $r_2=3.9228$, $q_1=1.972$, $q_2=2.968$, $a_{12}=-76.1573$ K, and $a_{21} = 438.005$ K. 
The saturated vapor pressures of pure ethanol and toluene are given by Antoine equation: $$\log_{10}\frac{P^\text{sat}}{\text{mmHg}} = A-\frac{B}{T/^\circ\text{C}+C}, $$
where $(A, B, C)$ for ethanol is $(8.11220, 1592.864, 226.184)$ and for toluene is $(6.95087, 1342.31, 219.187)$. These data come from Elliott and Lira \cite{elliott2012introductory}. The vapor is assumed to be an ideal gas mixture. With these information, the ground-truth phase diagram at $1$ atm ($760$ mmHg) is plotted in Fig. \ref{fig:phase_diagram}. 

\par
\begin{figure}[!t]
  \centering
  \includegraphics[width=\columnwidth]{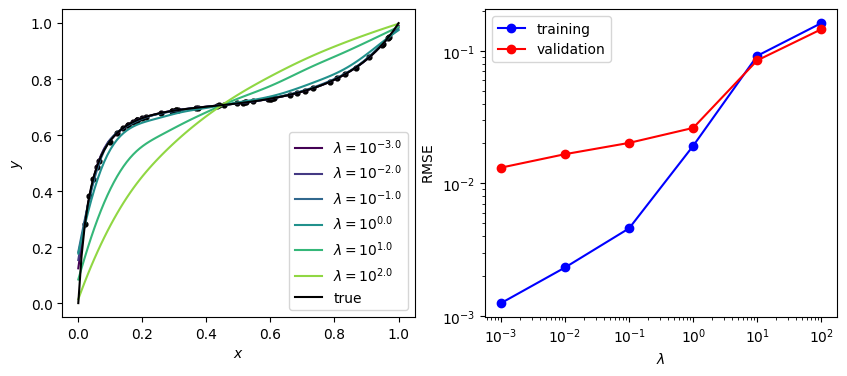}
  \caption{Vapor--liquid equilibrium curve identified through regularized regression near a reference model.}
  \label{fig:KRR}
\end{figure}

 Suppose a practitioner does not know of the true thermodynamic model (whose equations may be hidden in a simulator) but collected a sample of $50$ pairs of $(x, y)$ data. Out of an intuition based on Raoult's law (for ideal liquid mixtures), the user considers the following ``relative volatility'' model as the interpretable one:
$$h^\circ(x) = \frac{\alpha x}{\alpha x + (1-x)}, $$
where $\alpha$ is the relative volatility parameter, taken as a constant. By examining the ratio between the saturated pressures of ethanol and toluene, which approximately linearly varies with temperature (plot omitted), the user takes the average $2.973$ (between the normal boiling points of the two pure species) as a nominal value for $\alpha$. 
The discrepancy of the experimental data from the reference model $h^\circ$ is hence suggested to be identified by \eqref{eq:regression.1.KRR}, using a kernel function of $\kappa(x,x')=\exp(-100(x-x')^2)$. Using different regularization parameters $\lambda$, ranging from $10^{-3}$ to $10^2$, the resulting model $h$ is shown in Fig. \ref{fig:KRR} in comparison to the true model, from which the sample points are generated. 

\par In this case, the accurate model is recovered when $\lambda$ is chosen to be small enough values -- roughly, $\lambda<1$. 
Indeed, such a modeling approach, essentially the hybridization of a white-box fixed model and a black-box residual model, is able to arrive at a highly accurate relation. However, the fact that $\lambda$ cannot be chosen as larger values reflects the existence of a significant part in the $x$--$y$ relation that is ``uninterpretable''. 
This is clearly due to the low-quality reference model $h^\circ$ that the user assigned without a solid physical basis. Suppose that the user would like to consider the problem more seriously, a parameterized activity model that relates $\gamma_1, \gamma_2$ to $x$ and $T$ would be used, whose parameters would be identified along with the black-box residual.  This is our next problem setting.

\section{Setting II: Interpretable Subspace}
The user may have more than a single reference model as the physically interpretable model, if during the first-principles procedure, some parameters are left as unknown and to be determined. In such a setting, the consistency with first principles can be accounted for by a \emph{set of models}:
$$\mc{M} = \{ h_\theta(\cdot)= \phi(\cdot)^\top \theta: \theta\in \Theta \}.$$
Here $\phi(\cdot): \mathbb{X}\rightarrow \mR^N$ is a nonlinear mapping that transforms $x$ into some ``features''/``predictors'', and $\theta$ is a vector of parameters, belonging to a closed region $\Theta\subset \mR^N$. 
By this formulation, the set of interpretable models $\mc{M}$ is specified by a part of the $N$-dimensional function space $\{\phi(\cdot)^\top \theta: \theta\in \mR^N\}$. If each component mapping of $\phi$ is a member of an RKHS of infinite dimension, $\mc{H}$, then $\mc{M}$ is (a part of) a $N$-dimensional subspace of $\mc{H}$. 

\subsection{Formulation as Convex Regression}
The search for a hybrid model $h$ is therefore considered as optimization over $h = h_\theta + r$, where $h_\theta\in \mc{M}$ is a linearly parameterized interpretable model and $r\in \mc{H}$ is a black-box residual: 
\begin{equation}\label{eq:regression.2}
    \min_{\theta\in \Theta, \, r\in \mc{H}} \enspace \frac{1}{n}\sum_{i=1}^n \left( \phi(x_i)^\top \theta + r(x_i)-\eta_i \right)^2 + \lambda_\theta \|\theta\|^2 + \lambda_r \|r\|_{\mc{H}}^2 . 
\end{equation}
The regularization is performed on both $\theta\in \mR^N$ and $r\in \mc{H}$ to curb the model complexity. 
This problem is still a convex optimization problem (as long as $\Theta$ is a closed convex set with simple algebraic expressions). 

\par Since the finite-dimensional parameter space $\mR^N$ itself is naturally a RKHS and hence the optimization problem \eqref{eq:regression.2} is defined on $\mR^N \times \mc{H}$ (also an RKHS), according to the same representer theorem argument, the optimal residual model $r$ in the solution must be represented by a finite-dimensional kernel-based expression: $r(\cdot)=\sum_{i=1}^n c_i\kappa(x_i, \cdot)$. 
Then, by denoting
$$\Phi = \begin{bmatrix} \phi(x_1)^\top \\ \vdots \\ \phi(x_n)^\top \end{bmatrix}, \enspace \eta = \begin{bmatrix} \eta_1 \\ \vdots \\ \eta_n \end{bmatrix}, \enspace G = \begin{bmatrix} \kappa(x_1, x_1) & \cdots & \kappa(x_1, x_n) \\ \vdots & \ddots & \vdots \\ \kappa(x_n, x_1) & \cdots & \kappa(x_n, x_n) \end{bmatrix}, $$
the objective can be rewritten as 
$$ \min_{\theta\in \Theta, \, c\in \mR^n} \enspace \| \Phi\theta + Gc - \eta \|^2 + \lambda_\theta \|\theta\|^2 + \lambda_r c^\top Gc$$
which, if $\Theta = \mR^N$, admits an explicit solution:
$$ \begin{bmatrix} \theta^* \\ c^* \end{bmatrix} = \left( \begin{bmatrix} \Phi^\top \\ G^\top \end{bmatrix} \begin{bmatrix} \Phi&G \end{bmatrix} + \begin{bmatrix} \lambda_\theta I & 0 \\ 0 & \lambda_r G \end{bmatrix} \right)^{-1} \begin{bmatrix} \Phi^\top \\ G^\top \end{bmatrix} \eta. $$
We thus obtain the hybrid model:
$$h(x) = \phi(x)^\top \theta^* + \sum_{i=1}^n c_i^*\kappa(x_i, x). $$

\par Essentially, the difference between formulations \eqref{eq:regression.2} and \eqref{eq:regression.1} is that when using an interpretable subspace, $h_\theta$ (represented by $\theta$) and $r$ (represented by coefficients $c$) are simultaneously optimized. The hybrid model obtained can be better (with respect to the objective function in \eqref{eq:regression.2}) than the one obtained by fixing any $h_\theta \in \mc{M}$ as the reference model in \eqref{eq:regression.1}. 

\subsection{Example: Modeling Phase Equilibrium around Margules Model}
Seeking a hybrid vapor--liquid equilibrium model that weighs more on the physically interpretable part, the user considers using a parameterized activity coefficient model as the $h_\theta$, in addition to a black-box residual $r$. 
Suppose that the user is able to convert the phase diagram data (in the form of $T$--$x$--$y$) to the data of excess molar Gibbs energy, normalized by $RT$: 
$$ \frac{G^\text{ex}}{RT} = x\ln\frac{Py}{P^\text{sat}_1(T)} + (1-x)\ln\frac{P(1-y)}{P^\text{sat}_2(T)} .$$
The user thus decides to find a model that relates $G^\text{ex}/RT$ to $x$ as the argument. The class of interpretable models $\mc{M}$ comprises of the two-parameter Margules models: 
$$ G^\text{ex}/RT = \theta_1 x^2(1-x) + \theta_2x(1-x)^2, $$
where $\theta = (\theta_1,\theta_2) \in \Theta=\mR^2$. 

\par
\begin{figure}[!t]
  \centering
  \includegraphics[width=\columnwidth]{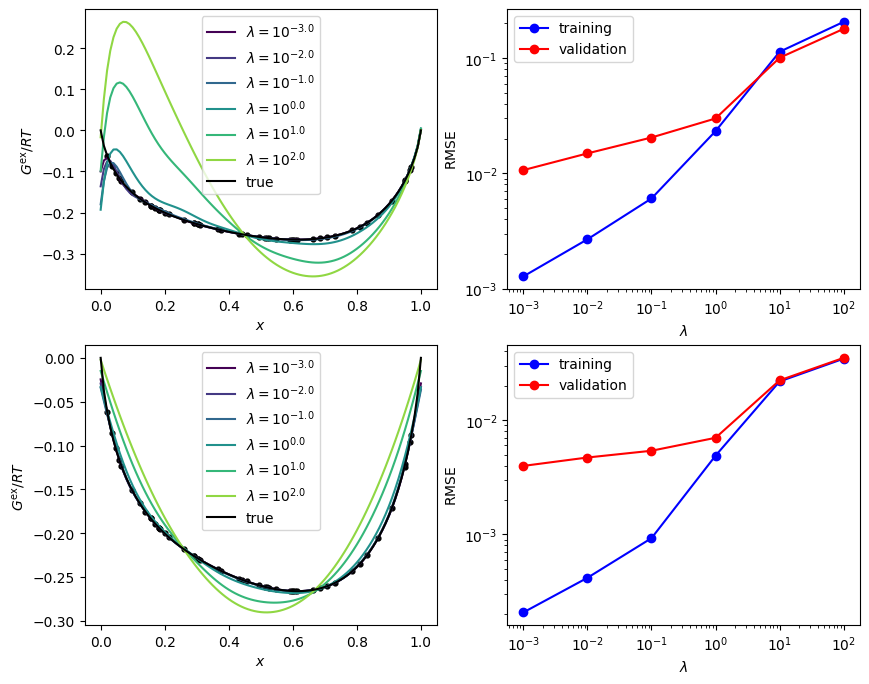}
  \caption{Excess Gibbs energy curve identified through regularized regression near a reference relative volatility model (upper) and near a Margules model class (lower).}
  \label{fig:KRR2}
\end{figure}

 By translating the user-defined reference model (based on a constant relativity) as a reference model on $G^\text{ex}/RT$ and repeating the procedure in the previous section, we obtain the results as shown in the upper half of Fig. \ref{fig:KRR2}, in comparison to the solutions to problem \eqref{eq:regression.2} using the Margules model class as the interpretable model subspace, shown in the lower half of Fig. \ref{fig:KRR2}. 
Obviously, the reference model largely deviates from the ground-truth excess Gibbs free energy, although on the $x$--$y$ curve, the deviation is not as much; actually, it is not even consistent with thermodynamics (as $x\rightarrow0$ or $1$, the excess Gibbs energy does not approach $0$). 
Comparing the two approaches, the later approach achieves much lower RMSE metric in both training and validation data under any given $\lambda$ value, indicating that the uninterpretable black-box residual model is significantly smaller. For example, at $\lambda=1$, the reference model approach does not generate a highly accurate model, while the later gives a hybrid model well aligned with data. 

\par However, in many process systems applications, the interpretable models embody physical first principles by a nonlinear parameterization structure: $h_\theta(\cdot) = h(\cdot|\theta)$, which generally cannot be represented in the form of $h_\theta(\cdot)=\phi(\cdot)^\top \theta$. 
For example, a mechanistic model of reaction network typically assumes that the comprising elementary reactions satisfy Arrhenius rate law, where the pre-exponential factor of the rate constant is proportional to $\exp(-E/RT)$ for some activation energy parameter $E$. The involvement of $E$ is clearly a nonlinear parameterization. 
For non-elementary reactions whose rates do not obey the law of mass actions, the rate laws cannot be reduced to a linear parameterization. Such cases are more complicated but should be more frequently encountered  in process systems, which are discussed in the next section.

\section{Setting III: Interpretable Manifold}
Now the interpretable class of models is $\mc{M} = \{h_\theta(\cdot)=h(\cdot|\theta): \theta\in \Theta\}$. There would not be a convex formulation similar to \eqref{eq:regression.2}. To enable a convex learning procedure, we seek a ``lifting'' approach -- such an idea is commonly used in the recent literature in control theory on Koopman operator. 
Specifically, on the input region $\mathbb{X}$ and the parameter region $\Theta$, we define two RKHS using kernels $\kappa$ and $\varkappa$, respectively. That is, $\mc{H}_x = \mc{H}_\kappa(\mathbb{X}) = \overline{\mr{span}}\{\kappa(x,\cdot): x\in \mathbb{X}\}$ and $\mc{H}_\theta = \mc{H}_\varkappa(\Theta) = \{\varkappa(\theta, \cdot): \theta\in\Theta\}$. These two spaces represent the functions of $x$ and functions of $\theta$ in scope, respectively. 

On these two infinite-dimensional RKHSs, the elements $\kappa(x, \cdot) = \phi(x) \in \mc{H}_x$ and $\varkappa(\theta, \cdot) = \varphi(\theta) \in \mc{H}_\theta$, respectively, are viewed as infinite-dimensional liftings, and hence representations, of points $x$ and $\theta$. 
We refer to $\phi(x)$ and $\varphi(\theta)$ as the ``canonical features'' in RKHSs. As such, the interpretable models form a manifold in the RKHS $\mc{H}_\theta$:
$$\mc{M} = \{\varphi(\theta): \theta\in \Theta\}.$$

\subsection{Tensor Product of RKHSs}
We introduce the following notion for the consideration of $h(\cdot|\cdot)$ as a function of both $x$ and $\theta$. 

\paragraph{Definition.} The tensor product of $\mc{H}_x$ and $\mc{H}_\theta$ is defined as the following Hilbert space, generated by products of functions of $x$ in $\mc{H}_x$ and functions of $\theta$ in $\mc{H}_\theta$:
$$\mc{H}_x\otimes \mc{H}_\theta = \overline{\mr{span}} \left\{ g_xg_\theta: g_x \in \mc{H}_x , \,  g_\theta \in \mc{H}_\theta \right\}$$
endowed with inner product, specified by:
$$\ip{g_xg_\theta}{h_xh_\theta} = \ip{g_x}{h_x}_{\mc{H}_x} \ip{g_\theta}{h_\theta}_{\mc{H}_\theta}$$
for any $g_x,h_x\in \mc{H}_x$ and $g_\theta,h_\theta\in \mc{H}_\theta$. 

\par For our hybrid modeling procedure, we shall therefore assume that any nonlinear parameterized model in consideration, $h(\cdot|\cdot)$, belongs to $\mc{H}_x\otimes \mc{H}_\theta$. Such a function space includes all functions of $x$ and $\theta$ that can be (implicitly) expressed as a (possibly infinite) linear combination of $\mc{H}_x$-class functions multiplied by $\mc{H}_\theta$-class functions. 
For example, if $h(\cdot|\cdot)$ is a polynomial in $x$ and $\theta$, and $\mc{H}_x$ and $\mc{H}_\theta$ contain all polynomials of $x$ and polynomials of $\theta$, respectively, then $h(\cdot|\cdot)$ indeed belongs to the tensor product space $\mc{H}_x\otimes \mc{H}_\theta$. If $h(\cdot|\cdot)$ is analytical and can be expressed as a Taylor series, then it is also included in the tensor product space. Hence, such a notion is clearly not restrictive, as physical laws are typically expressed as such analytical relations.\footnote{
For a more precise theory, the readers may be interested in knowing the equivalence between the RKHS specified by a Sobolev kernel and the Sobolev--Hilbert space of a certain smoothness index (e.g., Wendland \cite{wendland2004scattered}). The Sobolev--Hilbert space $W^{s,2}(\mathbb{X})$ refers to the class of functions on $\mathbb{X}$ that has square-integrable generalized derivatives up to $s$-th order. If $\mc{H}_x=W^{s,2}(\mathbb{X})$ and $\mc{H}_\theta = W^{r,2}(\Theta)$, then their tensor is the so-called anisotropic Sobolev space.}

\paragraph{Property.} $\mc{H}_x\otimes \mc{H}_\theta$ is a RKHS, whose kernel function is the product kernel $\kappa_\otimes: ((x, \theta), (x',\theta'))\mapsto \kappa(x,x')\varkappa(\theta, \theta')$. 
Thus, we denote $\kappa_\otimes((x,\theta), (\cdot, \cdot)) = \phi(x)\otimes \varphi(\theta)$, where $\otimes$ is a linear binary operation from $\mc{H}_x\times \mc{H}_\theta$ to $\mc{H}_x\otimes \mc{H}_\theta$, which is in fact the Kronecker product, generalized from finite dimensions to infinite dimensions.  

\par The property can be easily verified. Indeed, for any functions $g_x \in \mc{H}_x$ and $g_\theta\in \mc{H}_\theta$, due to the reproducing property on both spaces, we have $g_xg_\theta(x,\theta) = g_x(x)g_\theta(\theta) = \ip{g_x}{\kappa(x,\cdot)}\ip{g_\theta}{\varkappa(\theta,\cdot)} = \ip{g_xg_\theta}{\kappa(x,\cdot)\varkappa(x,\cdot)} = \ip{g_xg_\theta}{\kappa_\otimes((x,\theta), (\cdot,\cdot))}$. 

\paragraph{Corollary.} Given that the nonlinearly parameterized model $h(\cdot|\cdot)\in \mc{H}_x\otimes \mc{H}_\theta$, we have, for any $x\in \mathbb{X}$ and any $\theta\in\Theta$:
$$h(x|\theta) = \ip{h}{\phi(x)\otimes \varphi(\theta)}.$$
The nonlinear dependence of $h$ on model parameters $\theta$ is lifted into a  linear dependence on the canonical feature $\varphi(\theta)$ that lies on an infinite-dimensional RKHS. 

\subsection{Mixtures of Interpretable Models}
Now the set of all canonical features of admissible model parameters is $\Omega_0 = \{\omega=\varphi(\theta): \theta\in\Theta\}$, and hence $\mc{M} = \{ \ip{h}{\phi(\cdot)\otimes \omega}: \omega\in\Omega_0\}$. Since $\varphi: \Theta\to \mc{H}_\theta$ is nonlinear, $\Omega_0$ is a nonlinear manifold in the RKHS which is generally nonconvex. 
To the end of convex learning, we instead define the closed convex hull of canonical features of model parameters, $\Omega = \overline{\mr{conv}}(\Omega_0)$, namely: 
$$\Omega = \mr{cl}\left\{ \sum_{j=1}^m b_j\varphi(\theta_j): \begin{matrix}
    b_1, \cdots, b_m\geq 0, \, \theta_1, \cdots, \theta_m\in \Theta, \\
    b_1+\cdots+b_m = 1, \, m\in \mN
\end{matrix} \right\}. $$
Any element of $\Omega$ is now a convex combination of canonical features of a number of admissible parameters $\theta_1, \cdots, \theta_m$, or a limit of such convex combinations in the RKHS. 
In fact, $\Omega$ can also be expressed as 
$$\Omega = \left\{\int_\Theta \varphi(\theta)\xD{P}(\theta) : P\in \mc{P}(\Theta) \right\}$$
where $\mc{P}(\Theta)$ denotes the space of ``regular'' probability measures.\footnote{More accurately, $\mc{P}$ is the space of positive Borel measures that is normalized to an integral of $1$.}
The integral of canonical features under a probability measure $P$ is known as ``kernel mean embedding'' (KME) in the literature \cite{muandet2017kernel}. 

We thus redefine (actually enlarge) the class of interpretable models to be 
$$\mc{M} = \left\{ \ip{h}{\phi(x)\otimes \omega}: \omega\in\Omega \right\}. $$
Any element of this newly defined class $\mc{M}$ is interpreted as a ``stochastic mixture'' of nonlinearly parameterized models, i.e., a weighted average model under a multitude or a continuum of parameters. 
Now the nonlinear parameterization structure is resolved into a linear representation of the model by an infinite-dimensional parameter: $\omega\in \mc{H}_\theta$. 
Given an $\omega$ in the form of $\omega = \sum_{j=1}^m b_j\varphi(\theta_j)$, the model should be interpreted not only as an input--output relation $y=\sum_{j=1}^m b_jh(x|\theta_j)$, but also a predictive relation such that any $\theta$-dependent quantity, $q(\theta)=\ip{q}{\varphi(\theta)}$ (as long as $q\in \mc{H}_\theta$), should be predicted as $q=\sum_{j=1}^m b_jq(\theta_j)$. 

\par Now we can formulate the hybrid modeling problem as 
\begin{equation}\label{eq:KRR.3}
\begin{aligned}
    \min_{\omega\in\Omega, r\in \mc{H}_x} \enspace & \sum_{i=1}^n \left( \ip{h}{\phi(x_i)\otimes \omega} + \ip{r}{\phi(x_i)} - y_i \right)^2 \\
    &+ \lambda_\omega \|\omega\|_{\mc{H}_\theta}^2 + \lambda_r\|r\|_{\mc{H}_x}^2.
\end{aligned}
\end{equation}
The regularization terms are imposed on $\omega$ and $r$ on their RKHSs, respectively. The problem formulation remains convex. While the representer theorem applies partially to the choice of $r$, i.e., $r=\sum_{i=1}^n c_i\phi(x_i)$ must apply to the optimal solution of $r$, the optimal $\omega$ is not necessarily a finite convex combination. 
For an approximation, we sample $m$ points in $\Theta$: $\{\theta_1, \cdots, \theta_m\}$ and let $\omega = \sum_{j=1}^m b_j \varphi(\theta_j)$ for some $b_1, \cdots, b_m \geq 0$ with $b_1+\cdots+b_m=1$. Therefore, we reach at the following constrained quadratic programming problem:
\begin{equation}\label{eq:KRR.3.2}
\begin{aligned}
    \min_{b\in \mR^m, c\in \mR^n} \enspace & \|Hb + G_xc - y\|^2 + \lambda_\omega b^\top G_\theta b + \lambda_r c^\top G_xc \\
    \text{s.t.} \enspace & e^\top b=1, \enspace b\geq 0. 
\end{aligned}
\end{equation}
Here $G_x = [\kappa(x_i, x_{i'})]_{i,i'=1}^n$ is the Gram matrix formed by the sample points on $\mathbb{X}$, $G_\theta = [\varkappa(\theta_j, \theta_{j'})]_{j,j'=1}^m$ is the Gram matrix formed by parameter sample, and $H = [h(x_i, \theta_j)]_{1\leq i\leq n, 1\leq j\leq m}$ is the matrix of the model evaluated at sample points and sampled parameters. 

\par The hybrid model obtained by solving \eqref{eq:KRR.3.2} is therefore expressed as 
$$y = \sum_{j=1}^m b_j^* h(x|\theta_j) + \sum_{i=1}^n c_i^* \kappa(x_i, x). $$
In particular, if an ``optimal model parameter'' must be recommended, then, viewing $\theta\mapsto \theta$ as the identity function on $\Theta$, we get $\theta^*=\ip{\mr{id}}{\omega^*} = \ip{\mr{id}}{\sum_{j=1}^m b_j^*\varphi(\theta_j)}$, i.e., $\theta^*=\sum_{j=1}^m b_j^* \theta_j$. 

\subsection{Example: Modeling Phase Equilibrium around Wilson Model}
Seeking a more accurate model for valid--liquid equilibrium for ethanol--toluene binary mixture, the user considers Wilson model as the interpretable manifold:
$$\frac{G^\text{ex}}{RT}=-x_1\ln(x_1+x_2\Lambda_{12}) - x_2\ln(x_2+x_1\Lambda_{21}), $$
where
$$\Lambda_{12}=\frac{V_2}{V_1}\mr{e}^{-\frac{A_{12}}{RT}}, \enspace
\Lambda_{21}=\frac{V_1}{V_2}\mr{e}^{-\frac{A_{21}}{RT}}. $$
Wilson model was derived on a local composition theory and better accounts for polar, strongly non-ideal mixtures. 
The molar volumes for ethanol and toluene, $V_1$ and $V_2$, are taken as $58.7$ and $106.8$ mL/mol, respectively. 
We assume that $A_{12}$ and $A_{21}$ should lie in $[10^{-2}, 10^{2}]$, and hence consider $\theta_1 = (\log_{10} A_{12}+2)/4$ and $\theta_2=(\log_{10} A_{21}+2)/4$ as the two parameters on $[0,1]$. Following the discussed approach for model identification near an ``interpretable manifold'' (the nonlinearly parameterized Wilson model), we solve the problem \eqref{eq:KRR.3.2} via sampling on $\Theta = [0, 1]^2$ under a uniform distribution.  

\begin{figure*}[!t]
    \centering
    \includegraphics[width=\linewidth]{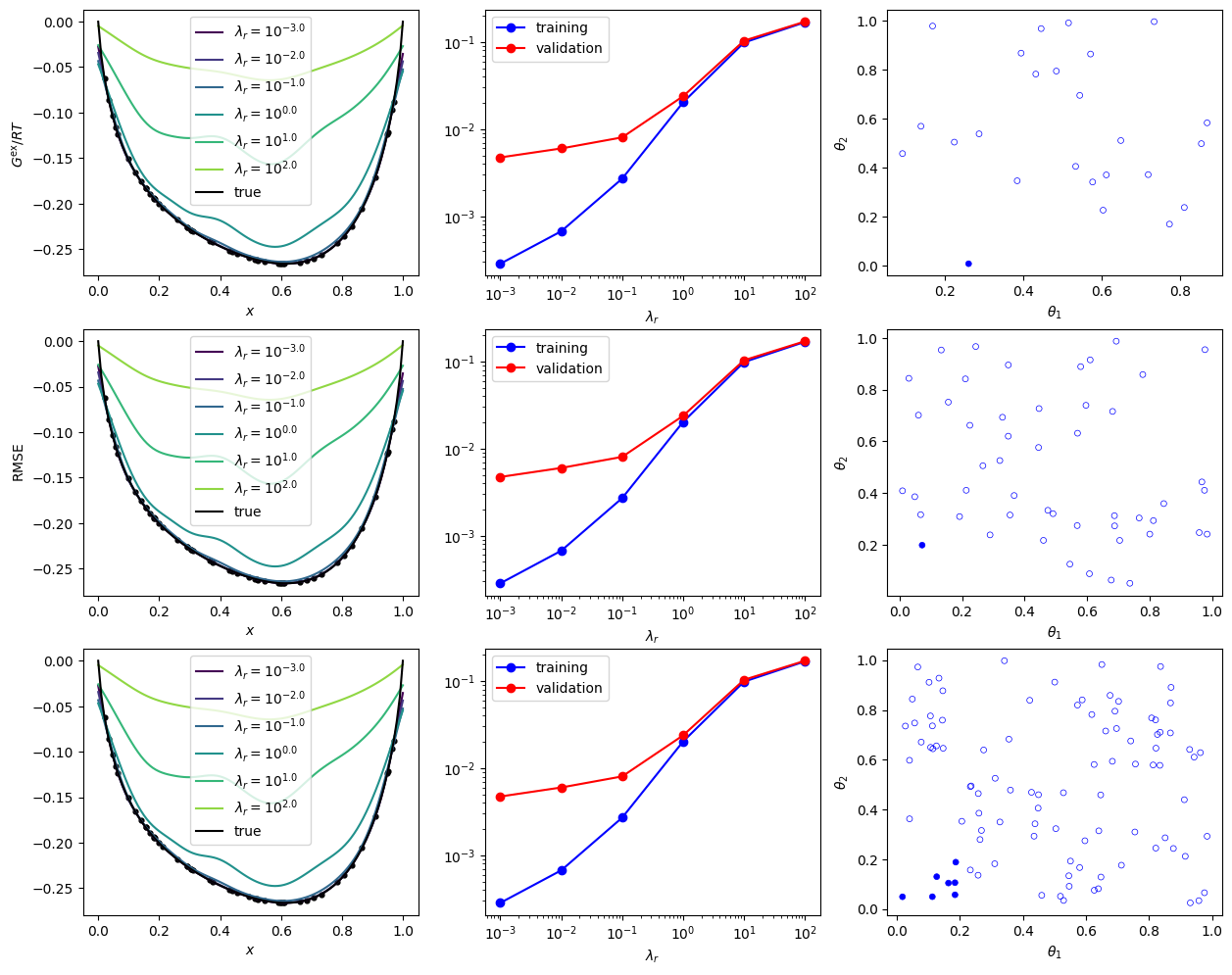}
    \caption{Excess Gibbs energy curve identified through regularized regression near the Wilson model ``manifold'' under different amount of sample points on the parameter space $\Theta$. (Upper: $m=25$, middle: $m=50$, lower: $m=100$.)}
    \label{fig:KRR3}
\end{figure*}
\par In Fig. \ref{fig:KRR3}, the results obtained under three sample sizes $m=25$, $m=50$, and $m=100$ on $\Theta$ are shown. In these experiments we keep $\lambda_\omega = 0$ (as there is no incentive to penalize the magnitude of $b$ in the presence of constraint $e^\top b = 1$) and use $\lambda=\lambda_r$ to regularize the residual model. 
Under these different $m$ values, the results appear almost identical. 
As seen in the third column of Fig. \ref{fig:KRR3}, where the scattered plots of sample points $\{\theta_j\}_{j=1}^m$ are shown and those points $\theta_j$ for which $b_j \geq 1/m$ are indicated by solid circles. The weights $b_j$ are concentrated on a small subset of $\{\theta_j\}$ points in the bottom-left corner, corresponding to small values of $A_{12}$ and $A_{21}$ Wilson parameters. 

Similar to the previous study on Margules model, the ``physically interpretable'' models are not sufficient to describe accurately the relation between excess Gibbs free energy and the binary mixture composition. Hence, the hybridization strategy is necessary, and the regularization coefficient plays the role of reconciliation between the first principles and empirical data.   
Comparatively, in the toluene--ethanol mixture, Wilson model does not accommodate as much regularization as Margules model; in particular, one can see that at $\lambda_r=10^0$, the error of hybrid Wilson is higher than that of hybrid Margules. Since both are two-parameter models, one can claim that for this mixture, Wilson model is not as physical as Margules, as it must use weaker regularization to allow the compensation of its inaccuracy by a black-box residual model.

\section{Extension to Dynamical Modeling}
Beyond static relations, the hybrid modeling approach using convex optimization can be extended to dynamical systems. 
Consider a continuous-time input-affine dynamical system, where $x(t)\in \mathbb{X}\subseteq \mR^{d_x}$ and $u(t)\in \mathbb{U}\subseteq \mR^{d_u}$ are referred to as states and inputs, respectively:
$$\dot{x}(t)=f(x(t), u(t)) = f_0(x(t)) + \sum_{k=1}^{d_u}u_k(t)f_k(x(t)). $$
For simplicity, in this paper, we assume that $f_k$ are known and $f_0$ is to be modeled (i.e., we know how inputs affect the states' velocity exactly, but are unsure about the detailed state dynamics). As such, we consider $f_0$ as an unknown part to be modeled for control. 
This assumption does not hurt generality, because even if all $f_k$ are unknown, we only need to augment the model to be identified without altering the nature of the formulation. 
In the recent control theory research, \emph{Koopman operator theory} has become a commonly used tool. We naturally extend the hybrid modeling approach in the previous section to dynamical modeling in the Koopman context.

\subsection{Preliminaries of Koopman Operator Theory}
\par Specifically, for an autonomous system $\dot{x}(t)=f(x(t))$, the Koopman semigroup refers to the family of operators, indexed by $t\geq 0$: 
$$[\mc{K}^t g](x) = g(F^t(x)) = [g\circ F^t](x),$$
where $F^t$ is the flow of the velocity field $f$. The Koopman semigroup is meaningfully defined, as a strongly continuous semigroup on a funtion space $\mc{G}$, if any $g\in \mc{G}$ guarantees $\mc{K}^tg\in \mc{G}$ and that $\lim_{t\downarrow0} (\mc{K}^tg-g) = 0$ pointwise.\footnote{
A typical choice of the function space $\mc{G}$ is the Sobolev space $W^{s,2}(\mathbb{X})$ as mentioned in Footnote 2. See, e.g., \cite{ye2026} for more details. In this paper, we take an ``elementary'' approach using an $N$-dimensional function space spanned by some basis functions and seek an approximate linear model thereon. A rigorous approach should explicitly assign an infinite-dimensional space $\mc{G}$ such as the Sobolev space.} 
With such a function space $\mc{G}$, $\lim_{t\downarrow 0} (\mc{K}^tg-g)/t = \mc{L}g$ defines an unbounded, closed, and densely defined linear operator $\mc{L}$, called the \emph{infinitesimal generator} of the Koopman semigroup. 
Its domain $\mr{dom}(\mc{L})$ is a dense subspace of $\mc{G}$ and contains the class of continuously differentiable functions $C^1(\mathbb{X})$, on which $\mc{L}g = \mr{D}g(x)f(x)$ (where $\mr{D}g(x)$ denotes the Jacobian matrix). 
As the system contains $d_u$ inputs instead of being autonomous, by defining all Koopman generators: $\mc{L}_0g(x) = \mr{D}g(x)f_0(x)$ and $\mc{L}_kg(x) = \mr{D}g(x)f_k(x)$ ($k=1,\cdots,d_u$), we get 
$$\xDD{g(x)}{t} = \mc{L}_0g(x) + \sum_{k=1}^{d_u} u_k\mc{L}_kg(x), \enspace \forall g\in C^1(\mathbb{X}). $$

\par The Koopman model can be identified via a least-squares routine called extended dynamic mode decomposition (EDMD) \cite{bevanda2021koopman}, whose original formulation for discrete-time systems. 
For a continuous-time system of form $\dot{x} = f(x)$, one can choose a finite amount of linearly independent basis functions of $\mc{H}$, all of which are continuously differentiable: $\psi=(\psi_1, \cdots, \psi_N)$, and solve the problem for data sample $\{x_i\}_{i=1}^n$ (which has an explicit solution):
$$\min_{A\in \mR^{N\times N}} \sum_{i=1}^n \| A\psi(x_i)-\dot{\psi}(x_i) \|^2.$$ 
This continuous-time version of EDMD is known as the \emph{generator EDMD} (gEDMD).\footnote{
The gEDMD formulation assumes that $\dot\psi(x_i)$ is available information, with $\dot{x}_i$ measurable. This is an unrealistic assumption, since the time derivative must be estimated and cannot be exact. Improved formulations based on the resolvent of Koopman semigroup has been proposed \cite{meng2026resolvent}. To avoid complexity, we still assume the availability of $\dot{x}$. 
}
Clearly, following this approach, the obtained solution for $A$ is a black-box estimation of the Koopman operator under the basis. Its implication is that for any given state-dependent function $g = \sum_{k=1}^N \gamma_k\psi_k$ that is in the span of the basis functions, at state $x$, the predicted time derivative is $\xD{g(x)}/\xD{t} \approx Ag(x) = \sum_{k=1}^N \gamma_kA\psi_k(x)$.

\subsection{Mixtures of Interpretable Koopman Models}
Under the setting that $f_k$ are all known while $f_0$ is unknown, the user should consider $\mc{L}_k$ as a known operator while $\mc{L}_0$ is to be determined. Suppose that by physical first principles, $f_0$ is parameterized nonlinearly as $f_0(\cdot|\theta)$ for $\theta\in\Theta$. Then the correspondingly Koopman generator $\mc{L}_0$ is nonlinearly parameterized: 
$$\mc{L}_0(\theta): g\mapsto \mr{D}g(\cdot) f_0(\cdot|\theta).$$
Thus, the goal is to identify $\mc{L}_0(\theta)$, in addition to a black-box ``residual Koopman generator'' $R$ that accounts for possible mismatch in the nonlinearly parameterized model structure. 
As we did in the previous section, the nonlinear dependence of $\mc{L}_0(\theta)$ on $\theta\in \Theta$ can be resolved by lifting $\theta$ into an infinite-dimensional RKHS $\mc{H}_\theta$ induced by a kernel function on $\Theta$, denoted as $\varkappa$ (and the corresponding canonical feature $\varphi$). 
Any model $\mc{L}_0(\theta)$ is thus reshaped as 
$$\mc{L}_0(\theta)g = \overline{\mc{L}} (g\otimes \varphi(\theta)) $$
with a linear mapping $\overline{\mc L}: \mr{dom}(\mc{L}_0) \otimes \mc{H}_\theta \subset \mc{G}\otimes \mc{H}_\theta \to \mc{G}$. 

The class of interpretable Koopman models is therefore expressed as the following class of stochastic mixtures of Koopman models:
$$\mc{M} = \left\{\mc{K}_\omega = g\mapsto \overline{\mc{L}}(g\otimes\omega):  \omega \in \Omega \right\}, $$
where
$$\Omega = \overline{\mr{conv}} \left\{ \varphi(\theta): \theta\in\Theta \right\}, $$
in other words, each ``interpretable model'' is represented by the average of a distribution of canonical features of parameters. Under $\omega = \int_{\Theta} \varphi(\theta)\xD{P(\theta)}$, given any $C^1$-function $g$, the predicted value for $\xD{g(x)}/\xD{t}$ is 
$$\overline{\mc L}(g\otimes \omega) = \int_{\Theta} \mr{D}g(\cdot) f_0(\cdot|\theta) \xD{P}(\theta). $$
In particular, given a (large) sample of points $\{\theta_j\}_{j=1}^m \subset \Theta$, and if $\omega = \sum_{j=1}^m b_j\varphi(\theta_j)$ (a finite linear combination), then the predicted $\xD{g\psi(x)}/\xD{t}$ for the given basis functions $\psi=(\psi_1,\cdots, \psi_N)$ is $\sum_{j=1}^m b_j\mr{D}\psi(\cdot) f_0(\cdot|\theta_j)$. 

\par In addition to the interpretable part $\mc{K}_\omega$, the hybrid part has yet an black-box residual, which therefore is an additional operator $\mc{K}_r$. This residual Koopman operator should be represented as a matrix $R\in \mR^{N\times N}$. 
Thus, the optimization problem for hybrid modeling is written as:
\begin{equation}\label{eq:KRR.dynamic}
    \begin{aligned}
    \min_{b\in \mR^m, R\in \mR^{N\times N}}  & \sum_{i=1}^n \left\| \sum_{j=1}^m b_j\mr{D}\psi\cdot f_0(x_i|\theta_j) + R\psi(x_i) - \dot{\psi}(x_i) \right\|^2 \\
    &\quad + \lambda_b\|b\|^2 + \lambda_R\|R\|_{\mr{F}}^2 \\
    \mr{s.t.} \enspace & b\geq 0, \enspace e^\top b=1. 
\end{aligned}
\end{equation}
Here $\|R\|_{\mr F}^2$ means the squared Frobenius norm of matrix $R$, namely the sum of squares of all its elements. 
By introducing short-hand notations $\psi_i = \psi(x_i)$, $\dot\psi_i = \dot\psi(x_i)$, $\dot{\Psi}_i = [\mr{D}\psi_k(x_i) f_0(x_i|\theta_j)]_{1\leq k \leq N, 1\leq j\leq m}$, $\mr{vec}\, R$ (column-major vectorization of $R$), and subsequently $M_i = \begin{bmatrix} \dot{\Psi}_i^\top \\ (\psi_i^\top \otimes I_N)^\top \end{bmatrix}$, we convert the problem into the following:
\begin{equation}\label{eq:KRR.dynamic.2}
    \begin{aligned}
    \min & \begin{bmatrix} b^\top & (\mr{vec}\,R) ^\top \end{bmatrix} \left( \sum_{i=1}^n M_iM_i^\top + \begin{bmatrix}
        \lambda_b I_m & 0 \\ 0 & \lambda_R I_{N^2}
    \end{bmatrix} \right) \begin{bmatrix} b \\ \mr{vec}\,R \end{bmatrix} \\
    &-2 \begin{bmatrix} b^\top & (\mr{vec}\,R) ^\top \end{bmatrix} \left( \sum_{i=1}^n M_i\psi_i \right) + \lambda_b\|b\|^2 + \lambda_R\|\mr{vec}\,R\|^2 \\
    \mr{s.t.} \enspace & b\geq 0, \enspace e^\top b=1. 
\end{aligned}
\end{equation}

By solving the above problem, a hybrid model is obtained for the autonomous part of the system. Adding the known input-affine part, the complete hybrid model is expressed as 
$$\dot{\psi}(x) \approx \sum_{j=1}^m b_j^\ast \mr{D}\psi(x)f_0(x|\theta_j) + \sum_{k=1}^{d_u} u_k\mr{D}\psi(x)f_k(x) + R^*\psi(x).$$
Yet, $\mr{D}\psi(x)f_0(x|\theta_j)$ ($j=1,\cdots,m$), as well as $\mr{D}\psi(x)f_k(x)$, may not be in the span of $\{\psi_1(x), \cdots, \psi_N(x)\}$. In other hands, when the state $x$ is lifted as the $N$-dimensional vector $\psi(x)$, the above model does not form a ``closed'' dynamics in $N$ dimensions. 
For simplicity and amenability to controller synthesis, we rather find an approximate closure: \footnote{These matrices $A_j$, $\Gamma_k$ and vectors $\beta_k$ can be obtained by any function approximation approach, e.g., least squares, and hence we omit their details for brevity. Because $f_0(x|\theta_j)$ and $f_k(x)$ are known functions of $x$, these approximations can be sought on new data points that do not need to come from a real process system but only come from a computer, and hence can be collected as many as points as possible.
Indeed, for a rigorous ``closure'', one must appeal to an infinite-dimensional state space such as the RKHS $\mc{H}_\kappa(\mathbb{X})$ specified by a well-chosen kernel (in particular, see \cite{tang-ye2025koopman}). Yet, the controller synthesis problem in infinite dimensions is difficult and such a Koopman-based synthesis problem remains open.}
\begin{equation}\label{eq:Koopman.closure}
    \mr{D}\psi(x)f_0(x|\theta_j) \approx A_j\psi(x), \enspace 
\mr{D}\psi(x)f_k(x) \approx \beta_k + \Gamma_k\psi(x). 
\end{equation}
As such, we obtain an approximate hybrid model, which is the following bilinear system on $N$ dimensions: 
$$\dot{\psi}(x)\approx \sum_{j=1}^m b_j^*A_{0j}\psi(x) + \sum_{k=1}^{d_u} u_k(\beta_k + \Gamma_k\psi(x)) + R^*\psi(x).$$
If we rewrite $\psi(x)=z \in \mR^N$, then
\begin{equation}\label{eq:Koopman.hybrid.final}
    \dot{z}\approx \left(\sum_{j=1}^m b_j^* A_{0j} + R^*\right) z + \sum_{k=1}^{d_u} u_k(\beta_k+\Gamma_kz). 
\end{equation}

\subsection{Example: Hybrid Koopman Modeling of Reactor Dynamics}
We consider the following continuously stirred tank reactor dynamics with a fractional rate law, obtained from Exercise Problem 7.7 of Kravaris and Kookos \cite{kravaris2021understanding}: $\dot{x} = f_0(x)+f_1(x)u$, $x(t)\in \mR^2$, $u(t)\in \mR^1$, in which
$$ f_0(x) = \begin{bmatrix}
    \frac{3-x_1}{4} - \frac{9(1+x_1)}{4(3+2x_1)} \\
    -\frac{3(1+x_2)}{4} + \frac{9(1+x_1)}{4(3+2x_1)}
\end{bmatrix}, \enspace 
f_1(x) = \begin{bmatrix}
    \frac{3-x_1}{4} \\
    -\frac{1+x_2}{4}
\end{bmatrix}.$$
Physically, $u$ is the throughput flow rate passing the tank, $x_1$ is the intermediate concentration, and $x_2$ is the product concentration, all in relative deviation ratios from their steady state values. We assume that $f_1$ (associated with convection dynamics) is known, while $f_0$ (containing kinetics is unknown), but parameterized nonlinear as 
$$f_0(x|\theta)=\begin{bmatrix}
    -\frac{1}{4}x_1-\theta_1x_1-\theta_2x_1^2 \\
    -\frac{3}{4}x_2+\theta_1x_1+\theta_2x_1^2
\end{bmatrix}$$ 
with parameters $\theta_1, \theta_2\in [0, 1]$. 
In the Koopman modeling, we choose the basis functions as 
$$\psi(x) = (x_1, \cdots, x_1^q, x_2, x_1x_2, \cdots, x_1^{q-1}x_2), $$ 
including monomials of $x_1$ up to maximum degree $q$ and these monomials up to degree $q-1$ multiplied by $x_2$. Thus $N=2q$. Without nonlinearities in $x_2$, it can actually be justified that $\mr{D}\psi(x)f_1(x) = \beta + \Gamma\psi(x)$ is exact (instead of ``$\approx$''). 
In this experiment, we fix $q=3$. 

\par
\begin{figure}[!t]
    \centering
    \includegraphics[width=\columnwidth]{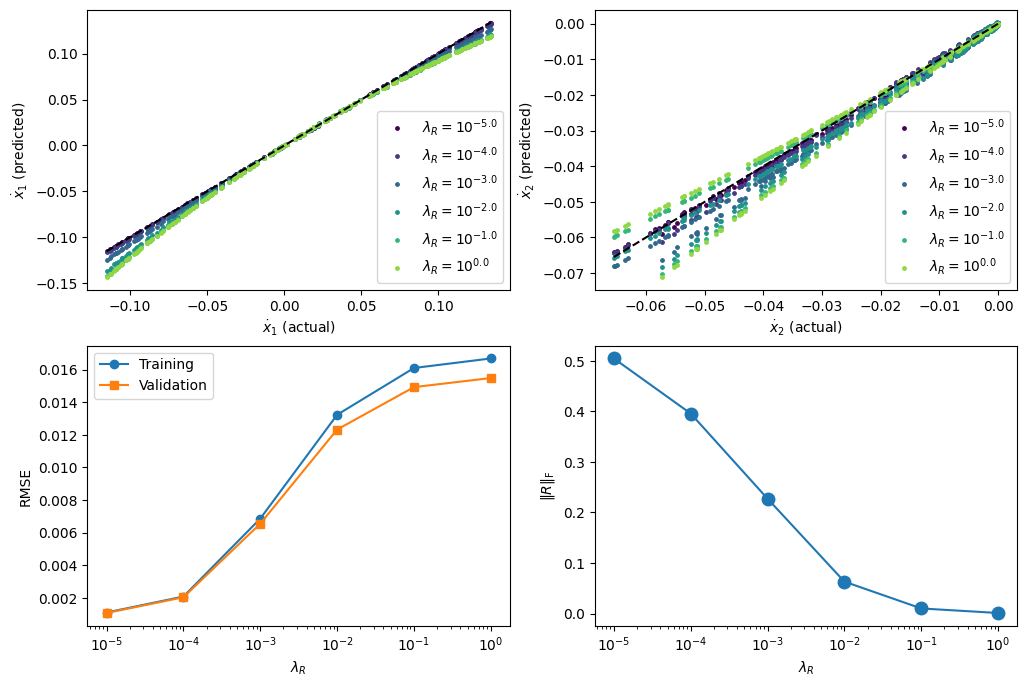}
    \caption{Prediction performance of the hybrid Koopman models identified under difference choices of the regularization coefficient $\lambda_R$. Upper two subfigures: predicted $\dot{x}$ versus actual $\dot{x}$. Lower left subfigure: training and validation errors. Lower right subfigure: $\|R\|_{\mr{F}}$ values.}
    \label{fig:KRR_dyn}
\end{figure}

 Using $n=200$ sample points over the state space $\mathbb{X} = [-1/4, 1/4]^2$ and $m=25$ sample points on the parameter space $\Theta = [0, 1]^2$, we obtain $m$ matrices $A_{j} \in \mR^{N\times N}$ as well as $(\beta, \Gamma)$ in \eqref{eq:Koopman.closure} via least squares. 
By solving $b^\ast$ and $R^\ast$, we arrive at the final hybrid Koopman model \eqref{eq:Koopman.hybrid.final}. The total number of degrees of freedom is the number of entries in $R$ plus the dimension of $b$, namely $4q^2 + m = 61$. 
Under a range of choices for the regularization coefficient $\lambda_R$ in \eqref{eq:KRR.dynamic.2}, the learning outcomes are shown in Fig. \ref{fig:KRR_dyn}. 
As expected, as $\lambda_R$ increases, the black-box residual operator (approximate) $R$ is more strongly penalized and hence must be smaller, and hence the predicted $(\dot{x}_1, \dot{x}_2)$ (which is extracted from the first two components of the right-hand side of \eqref{eq:Koopman.hybrid.final} at every $x$) must deviate from the actual values of $(\dot{x}_1, \dot{x}_2)$. 
The training error and validation error (obtained on resampled new data) must increase with increasing $\lambda$, and $\|R\|_{\mr{F}}$ decreases. 

\par
\begin{figure}[!t]
    \centering
    \includegraphics[width=0.5\textwidth]{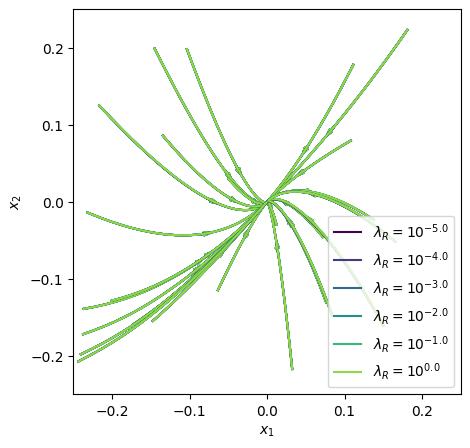}
    \caption{Trajectories under Lin--Sontag controllers using the ground-truth model and hybrid Koopman models.}
    \label{fig:control}
\end{figure}

 Finally the learned hybrid Koopman models are used in a Lyapunov-based controller. 
We consider $V(x) = \|\psi(x)\|^2 = (x_1^2 + x_2^2)(1 + \cdots + x_1^{2(q-1)}) = \|x\|^2(1-x_1^{2q})/(1-x_1^2)$ as the control Lyapunov function, and force $V(x)$ to decrease according to the Lin--Sontag formula \cite{lin1991universal} to design a controller under constraint $|u|\leq 1$.  
The comparison of trajectories, issued from some randomly chosen initial states on $\mathbb{X}$, under the ground-truth model-based controller and under the hybrid Koopman model-based controllers, are shown in Fig. \ref{fig:control}. These trajectories are almost overlapping, even at large $\lambda_R$ where the resulting residual operator $R$ becomes very small. 
This indicates that for retaining the control performance of the ideal true model, the convex identification of a ``mixture'' of nonlinearly parameterized model suffices to provide a satisfactory approximate Koopman operator. 


\section{Conclusions}
In this paper, focusing on the problem of hybrid modeling for process systems, a convex learning approach is proposed. Whether the physical interpretability or first-principles knowledge is accounted for by (i) a single reference model, (ii) a linear parameterization, or (iii) a nonlinear parameterization structure, the hybrid model, comprising of a white/gray-box interpretable part and a black-box residual part, can be learned via convex optimization. 
The approaches here proposed are based on the calculus of reproducing kernel Hilbert spaces (RKHS), which provides a general and systematic tool for nonlinear data-driven modeling. 
It is also shown in this paper that the convex learning approach can be applied to not only static systems but also dynamical ones, in which the ``model'' to be identified is an operator, namely the (infinitesimal generator) of the Koopman semigroup in continuous time or Koopman operator in discrete time. 

\par With the convex learning problems being not more difficult than quadratic programs, the learning procedure becomes intrinsically more reliable and computationally efficient. 
The resulting hybrid models, expressed as a fixed or first-principles parameterized model plus a combination of kernel functions, appear much simpler than a neural network or decision tree surrogate, which tends to be overparameterized and can be difficult to be embedded in a decision-making optimization or control routine. 
While many recent efforts aimed to achieve optimization under neural network or decision tree models \cite{tsay2021partition, ceccon2022omlt, misener2023formulating}), it would still be beneficial if the model structure can be made simpler during the modeling phase. 

\par For process control, where the feedback control actions must be computed in real time, the research on nonlinear systems increasingly relies on Koopman operator theory that promises to solve control problems in globally linearized ways. 
Here, we showed that the hybrid Koopman operator modeling is beneficial for learning an accurate Koopman model near a physically interpretable nonlinear model, thus restoring the Lyapunov-based control performance. 
Hence, there is no non-convexity involved from modeling to control, enabling end-to-end convex computation. This approach is in sharp contrast to neural-based Koopman modeling \cite{lusch2018deep} or typical Koopman modeling on function spaces where the operator model is completely black-box.

\section*{Acknowledgments}
The work is supported by NSF CBET \#2414369. The codes for this work are available at \url{https://github.com/WentaoTang-Pack/ConvexHybridModeling}. 
\par During the peer review process for one of the author's papers, an anonymous reviewer asked the question of ``interpretability of Koopman modeling''. This directly motivated the author's development of the present paper. Discussions with Profs. Yankai Cao and Ilias Mitrai during past AIChE annual meetings also gave the author a strong feeling that the concept of convexity should be instilled into the research on ``interpretable/explainable AI''. 

\newpage
\bibliographystyle{ieeetr}
\bibliography{mybib.bib}
\end{document}